\documentclass[a4paper,10pt]{iopart}

\usepackage{graphicx}

\begin{document}

\title[Determining the phonon DOS from specific heat measurements ...]{Determining the phonon DOS from
specific heat measurements via maximum entropy methods}

\author{J.P.Hague} 

\address{Department of Physics and Astronomy, University of Leicester,
Leicester}

\begin{abstract}
The maximum entropy and reverse Monte-Carlo methods are applied to the
computation of the phonon density of states (DOS) from heat capacity
data. The approach is introduced and the formalism is
described. Simulated data is used to test the method, and its
sensitivity to noise. Heat capacity measurements from diamond are used
to demonstrate the use of the method with experimental
data. Comparison between maximum entropy and reverse Monte-Carlo
results shows the form of the entropy used here is correct, and that
results are stable and reliable. Major features of the DOS are picked
out, and acoustic and optical phonons can be treated with the same
approach. The treatment set out in this paper provides a
cost-effective and reliable method for studies of the phonon
properties of materials. [Published as J. Phys: Condens. Matter. {\bf 17}, pp2397-2405 (2005)]
\end{abstract}

\section{Introduction}

It is nearly a century since Einstein's paper ``\emph{The Planck
theory of radiation and the theory of specific heat}'' was published
\cite{einstein1907a}. This well cited paper detailed a simple
approximation for the phonon contribution to the specific heat of
solids. Along with Debye's theory of the specific heat due to acoustic
phonons \cite{ashcroft} it has become the stuff of textbook legends. Both
methods are still used regularly (e.g. to separate the magnetic and
phonon contributions to the specific heat \cite{chung2004a}).

Interest in lattice vibrations is timely because of the discovery that
phonons play a significant role in the physics of cuprates
\cite{lanzara2001a}. In this regard, it would be useful to have
greater experimental access to information about phonons. Currently
there are two major methods used to probe lattice excitations; Raman
(and Raleigh) scattering, where only the zone-centre phonons are
measured \cite{ashcroft} and neutron scattering, which requires large
scale facilities. Even with recent neutron scattering developments
such as MAPS at ISIS, the observation of phonon modes across the
entire Brillouin zone is extremely difficult, and an array of large
single-crystals is required. In a typical triple-axis instrument, only
a few points along high symmetry directions are determined, with the
phonon DOS inferred from a model of internuclear forces using the
dynamical matrix formalism \cite{ashcroft}. There have been some
recent attempts to classify the phonon DOS from specific heat
measurements \cite{loram,wen}. The approach in reference \cite{loram}
was relatively effective and could be applied to experimental data,
but suffered from unphysical negative densities of states. Also, the
basis set decomposition that was described is expected to be
significantly less effective when applied to systems with both
acoustic and optical phonons.

The method described in this paper may be used to classify the phonons
in many different kinds of materials, and has the advantage that the
only equipment requirement is a heat-capacity rig and a modern
personal computer. Both these pieces of equipment are very common in
solid state laboratories and therefore I expect this method to be of
great use to the condensed matter community. The paper is organised as
follows: The formalism of the maximum entropy and reverse Monte-Carlo
methods is introduced in section \ref{sec:formalism}. In section
\ref{sec:results}, I test the method using simulated heat capacity
data. I also determine the phonon DOS of diamond. Finally conclusions
and recommendations are given in section \ref{sec:conclusions}.

\section{Methodology and formalism}
\label{sec:formalism}

The maximum entropy method (MAXENT) is based on the principles of
Baysian inference and can be used to provide a general framework for
the solution of inverse problems in physics. In this section, I
describe how maximum entropy methods can be used to extract the
density of phonon states from heat-capacity data.

The specific heat due to phonons is related to the phonon density of
states according to the integral transform,
%
% check the prefactors here
\begin{equation}
c_v(T)= 3R \int_0^{\infty} d\omega\, D(\omega) \frac{\hbar^2\omega^2
e^{\hbar\omega/kT}}{(kT)^2(1-e^{\hbar\omega/kT})^2}
\end{equation}
%
% Kernel may be rewritten as 1/sinh^2(w/2T)
%
where $D(\omega)$ is the phonon DOS and R is the gas constant. For
non-magnetic insulators, this is the major form of the specific
heat. The DOS may be discretised by writing $D(n\Delta\omega)\equiv
s_n\Delta\omega$, so that the problem is reformulated as a matrix
equation, $c_v(T_i)=\sum_j K_{ij} s_j$. The kernel matrix is written
as $K_{ij}=3Ry^2 e^y/(1-e^y)^2$ and $y=\hbar\omega/kT$. The matrix may
not simply be inverted, since there are typically more unknowns than
data points, leading to a serious over-fitting of the data. While the
integral transform can be inverted, but the problem is ill conditioned
it is very sensitive to errors in the data. In particular,
experimental noise and discrete (incomplete) data will negate the
results \cite{montroll,kroll,chambers}. Some attempts have been made
to ease the problems of ill conditioning by using alternative basis
sets \cite{loram,wen}. However, these lead to negative densities of
states, and in particular, the results of reference \cite{wen} have a
clearly unphysical form for low frequency phonons, which, being
acoustic in nature should tend to zero according to an $\omega^2$ form
rather than diverging. Similarly, one may not simply apply least
squares fitting, since for useful results, there are significantly
more fitting parameters than data points. A common approach to
avoiding over-fitting involves introducing a regularization
process. Two schemes will be discussed in this paper: The maximum
entropy scheme and the reverse (or Markov chain) Monte-Carlo method.

The essence of the MAXENT approach is that the most probable fit to an
inverse problem in the absence of data is located where a large number
of similar data configurations are present. A large number of similar
configurations corresponds to a minimally contorted spectrum (or DOS
in this case). Small changes in the DOS then lead to small changes in
the value of $\chi^2$. In the current problem, over-fitting would lead
to a $D(\epsilon)$ that rapidly oscillates from negative to positive
values, which is clearly unphysical since the DOS must be
positive. Formally, the regularisation is introduced via a ``free
energy'', which is a sum of the familiar $\chi^2$ goodness of fit
parameter and an entropy term, $S$,
\begin{equation}
\mathcal{F}=\chi^2\{s_i\}+\alpha S\{ s_i\}
\label{eqn:freeenergy}
\end{equation}
where,
\begin{equation}
\chi^2=\sum_{i=1}^{N_d}\frac{c_{\mathrm{obs}}(T_i)-c_{\mathrm{calc}}(T_i)}{\sigma^2_i}
\end{equation}
\begin{equation}
S=\sum_{i=1}^{N_s} [ m_i -s_i + s_i \ln (\frac{s_i}{m_i}) ]
\end{equation}
$\sigma^2$ is the standard deviation, $N_d$ is the number of data, and
$N_s$ is the number of discrete points in the DOS.

The entropy is multiplied by a hyper-parameter (nuisance parameter),
$\alpha$, which has a similar role to temperature in the analogous
statistical system. An additional spectrum, $m_i$ is introduced, which
is known as the default model. The default model contains information
that is already known about the system, and is the default result of
minimising the free energy (equation \ref{eqn:freeenergy}) in the
absence of data, i.e. without the $\chi^2$ term. Since in this problem
I examine the density of states, then it is known that the spectrum,
$s_i$, may only contain positive values. It is also known that the
normalisation of the spectrum is the number of atoms in the unit
cell. Therefore, the default model chosen throughout this work is a
flat positive distribution normalised to the number of modes. Nothing
else is assumed.

The maximum entropy method comes in several different flavours. In
this paper, I use Bryan's algorithm \cite{bryan1989a}. Bryan's
algorithm differs somewhat from the Historic approach since the
hyper-parameter $\alpha$ is chosen from a continuous probability. In
historic MAXENT, $\alpha$ is typically chosen so that
$\chi^2=N_d$. This is intuitive, but will normally lead to
under-fitting of the data, since neighbouring data points can have
closely related values. Closely related data points result in an
effective error of the combined points which is lower that that of
individual points considered separately.

When implementing Bryan's algorithm, the spectrum is calculated from
the weighted average of spectra calculated for all $\alpha$ values,
\begin{equation}
s_i=\int_{0}^{\infty}P[\alpha|s_{\alpha}]s_{i\alpha}
\end{equation}
The probability distribution $P[\alpha|s]$ is calculated as in
reference \cite{bryan1989a}. A singular value decomposition is also
introduced to reduce the total number of search directions. The
resulting algorithm is fast and with the integration over the nuisance
parameter is fully consistent with Baysian analysis. Furthermore, a
positive density of states is guaranteed.

The reverse Monte-Carlo (RMC) method is used to validate the MAXENT
results and to demonstrate an alternative scheme. The RMC method
treats all possible data configurations as an ensemble, and averages
over all possible data sets, weighted by the likelyhood function
$e^{-\chi^2/2T}$. A new parameter, $T$ is introduced, which is a
nuisance parameter similar to $\alpha$ \footnote{Note that $T$ acts as
a temperature, but is not the same as the temperature argument of the
specific heat}. In the spirit of statistical mechanics, the Metropolis
algorithm with $E\equiv\chi^2$ is used here as an efficient approach
to the averaging of the spectrum over the ensemble defined by the
likelyhood.

The algorithm proceeds as follows: A change to a variable is
suggested. If the change leads to a reduction in $\chi^2$, then it is
accepted. Otherwise, it may still be accepted according to the
probability $P=\exp(-\Delta E/kT)$ ($\Delta E$ is the difference
between $\chi^2$ values before and after the change). Such a scheme is
one of the simplest which is consistent with the principle of detailed
balance. It is instructive to note the close relationship between
reverse Monte-Carlo and MAXENT algorithms. In the event of uncontorted
data, there will be many available states for the RMC algorithm, and
the overall sampling rate is greatly increased. Put another way, the
local entropy of the configuration space is higher. In this way,
states with higher entropy are favoured, and the conceptual similarity
between the methods can be seen \footnote{If the local entropy could
be measured in the RMC method, then it would be possible to show that
RMC and MAXENT are fundamentally equivalent}. The advantage of the RMC
algorithm is that no form for the entropy is assumed, and it is
conceptually very simple. However, it is computationally intensive,
and as I demonstrate, the spectra resulting from MAXENT and RMC are
essentially identical.

A few modifications to the algorithm are made for the current
application. To reduce the parameter space, the spectrum is
constrained to be normalised to $N_{\mathrm{ph}}$. In practical terms,
this means that for every change $s_i\rightarrow s_i+r_i$, there is an
equivalent $s_j\rightarrow s_j-r_j$ with $i\neq j$. I will term this
spectral RMC. The spectrum is also constrained to positivity, so any
updates that violate that condition for either of $s_i$ or $s_j$ are
discarded. The calculation of the difference in $\chi^2$ which scales
as $\mathcal{O}(N_dN_s)$ can be rewritten as an $\mathcal{O}(N_d)$
process when only one variable in the spectrum is changed leading to a
massive increase in speed. The energy landscape of $\chi^2$ is
complicated and has many troughs, some of which may be deep. To ensure
that all troughs are sampled, $r_i$ is chosen from a random variate
obeying the Cauchy distribution,
\begin{equation}
P(r_i)=\frac{\sigma_i}{\sigma_i^2+r_i^2}
\end{equation}
The Cauchy distribution has been widely applied to fast simulated
annealing, and is designed to cover the parameter space quickly. It is
also ideal for Monte-Carlo simulations with continuous variables where
the temperature is held fixed while expectation values of variables
are taken. The width of the distribution, $\sigma_i$ is changed every
few iterations to keep the acceptance rate close to $70\%$. This is
essential for a fast computation (otherwise spectral points with a
larger magnitude are favoured in the update).

To initialise the algorithm, the update scheme is run for a few
hundred thousand iterations until thermal equilibrium is reached. Data
are measured using a blocking scheme, where the blocking size is much
greater than the correlation time. In this way a reliable estimate of
the error on each point can be obtained. Finally, the algorithm is
stopped when the data error is approximately $1\%$. Since the data
error scales as $1/\sqrt{N}$, this is the best accuracy achievable on
a modern workstation in a few hours of calculation.

\section{Results}
\label{sec:results}

In this section, I present results showing the determination of the
phonon DOS from a simulated specific heat measurement. The densities
of states that are used here are simpler in form than those expected
in a real solid. However, they contain the sort of features that might
be expected in real systems, such as narrow high-energy optical phonon
modes, discontinuities associated with Brillouin-zone boundaries and
the typical low-energy $\omega^2$ behaviour associated with acoustic
phonons, and therefore constitute a fair test of the method.

\begin{figure}
\begin{indented}\item[]
\includegraphics[width=70mm,height=100mm,angle=270]{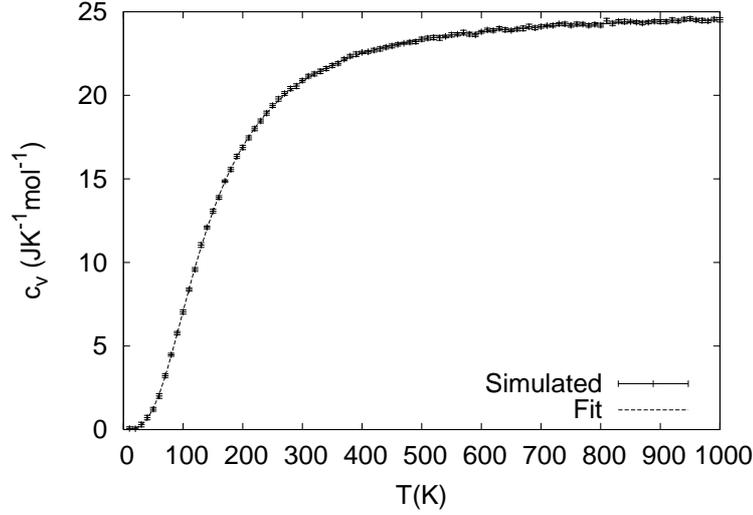}
\end{indented}
\caption{Simulated data corresponding to a Debye spectrum with
$\omega_D=50$meV. An error of 0.3\% has been added. Also shown is the
fit to the data using Bryan's algorithm. The Debye spectrum has the
correct low frequency behaviour and also has a sharp cutoff typical of
the phonon DOS at the Brillouin zone boundary.}
\label{fig:simdata}
\end{figure}

An example of simulated data for a Debye spectrum with
$\omega_D=50\mathrm{meV}$ is shown in figure \ref{fig:simdata}. Also
shown is the fit to the data from the MAXENT method. Gaussian noise is
added to the data, and several ``data sets'' are calculated. The
average and RMS values are then computed to determine the mean and
variance, from which the error can be calculated. A similar approach
to calculating the error should be used when carrying out an
experiment. For a better estimate of the $\chi^2$ value, the
covariance matrix may be determined, and will lead to improvements in
the results. This is unnecessary for simulated data where the noise
added to neighbouring data points is statistically independent, and
off diagonal terms in the covariance vanish. Simulated data points
were calculated to cover the full temperature range from low
temperatures ($T\ll 50\mathrm{meV}$) to high temperatures approaching
the saturation associated with the Dulong--Petit law. The data error
in this case is $0.3\%$ of the saturation value, or $\sim 0.075
\mathrm{JK}^{-1}\mathrm{mol}^{-1}$. Such an error is easily achievable
with standard laboratory equipment.

\begin{figure}
\begin{indented}\item[]
\includegraphics[width=70mm,height=100mm,angle=270]{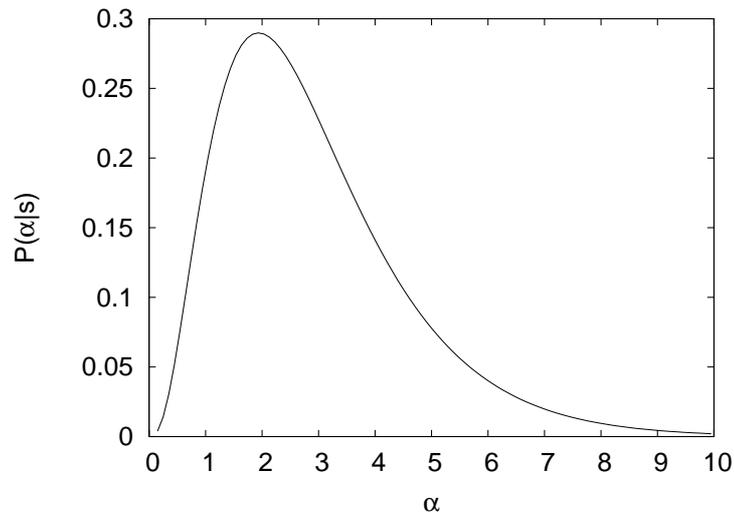}
\end{indented}
\caption{Alpha probability for Bryan's algorithm (left axis). Note
that the distribution is not sharp as is assumed in the case of
classic and historic MAXENT. Calculations are carried out for a series
of $\alpha$ values, and then a weighted sum of resulting spectra is
computed to arrive at the most probable DOS. The input data was a
Debye specific heat with $0.3\%$ data error as shown in \ref{fig:simdata}.}
\label{fig:probalpha}
\end{figure}

It is instructive to note that the integration over the nuisance
parameter inherent in Bryan's algorithm is important. The distribution
of $P[\alpha|s]$ is shown in figure \ref{fig:probalpha}. Note that the
distribution is not a sharply peaked $\delta$-function as is assumed
in the case of classic and historic MAXENT. Clearly the distribution
is significant for $0<\alpha<12.5$. A weighted sum of the resulting
spectra is therefore needed to arrive at the most probable DOS, and
historic and classic methods which treat only one $\alpha$ value
should be approached with care.

\begin{figure}
\begin{indented}\item[]
\includegraphics[width=70mm,height=100mm,angle=270]{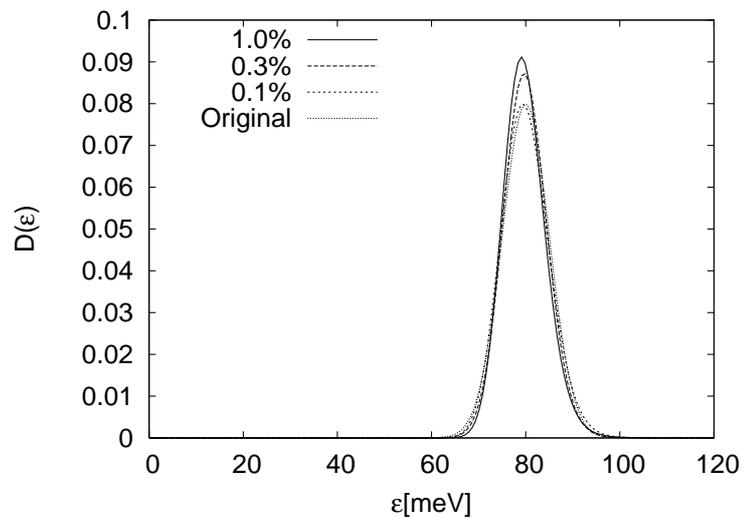}
\end{indented}
\caption{Simulated phonon DOS representing a broadened ``optical''
phonon, and the spectra extracted using Bryan's algorithm. The
simulated density of states is a Gaussian centred about 80meV with a
full-width half-maximum (FWHM) of 5 meV normalised to unity. The
method performs well, and is accurate for an achievable data
error. 100 simulated data points covering a complete temperature range
from low temperatures ($T \ll 80\mathrm{meV}$) to high temperatures approaching
the Dulong--Petit law. The simulated spectrum was forward transformed,
and then Gaussian noise was applied before inverting the transform.}
\label{fig:optical}
\end{figure}

A very simple model is shown in figure \ref{fig:optical}. As before,
the simulated spectrum was forward transformed to determine $c_v(T)$,
and then Gaussian noise was applied before inverting the
transform. The simulation technique ensures that there is an exact
answer to compare with. The underlying DOS is a broadened ``optical''
phonon represented as a Gaussian, and spectra computed using Bryan's
algorithm are shown for various data error. The data error is a
percentage of the Dulong--Petit saturation value. The method performs
well with all spectra having a reasonable agreement. As expected, the
method performs better for more accurate data. The simulated data
error shown here is achievable using standard laboratory equipment.

\begin{figure}
\begin{indented}\item[]
\includegraphics[width=70mm,height=100mm,angle=270]{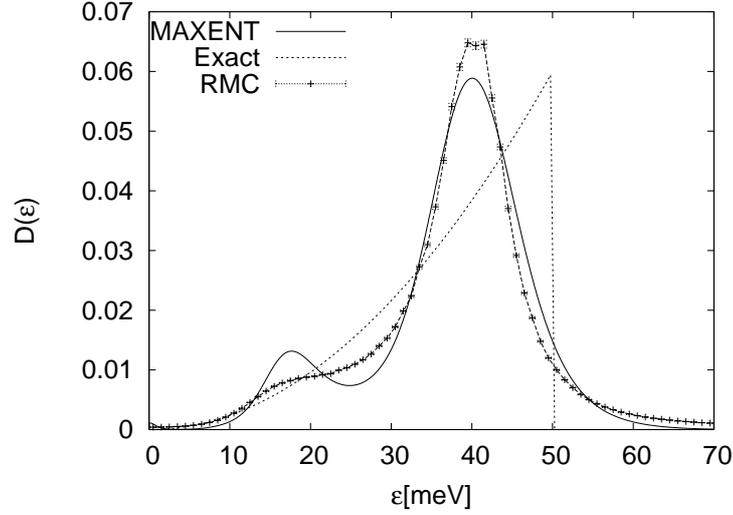}
\end{indented}
\caption{Simulated phonon DOS representing an acoustic phonon. The
same approach as in fig. \ref{fig:optical} was used to simulate the
data. Bryan's algorithm was less successful in this case because the
discontinuity in the phonon DOS leads to ringing. This suggests that
decomposition into a non-Gaussian basis might be more
appropriate. Also shown is the result from a reverse Monte-Carlo
calculation. RMC is quite general, however, calculations take a few
hours, in contrast to a few seconds for the MAXENT algorithm. The RMC
algorithm suffered from less ringing, however anomalous high-frequency
states were present. In both cases, the general features of the
spectrum are recovered. It can be seen that as few as 100 data points
are necessary to determine the spectrum. The accuracy used here was
$0.3\%$.}
\label{fig:acoustic}
\end{figure}

In order to simulate data consistent with acoustic phonons, the simple
Debye model was used ($D(\epsilon)=3\omega^2/\omega_D^3$). Although
this is a crude DOS, it has many of the features of the acoustic
phonons, including the $\omega^2$ behaviour at low frequencies, and a
sharp cutoff consistent with the effects of the zone boundary,
i.e. the cutoff can be thought of as representing a van Hove
singularity. Results for the analysis are shown in figure
\ref{fig:acoustic}. The low frequency behaviour is recovered, however
Bryan's algorithm was less successful for the high frequency behaviour
because the discontinuity in the phonon DOS leads to ringing. This
suggests that decomposition into a non-Gaussian basis might be more
appropriate. Also shown is the result from a reverse Monte-Carlo
calculation. RMC is quite general, however, calculations take a few
hours, in contrast to a few seconds for the MAXENT algorithm. I use a
Cauchy scheme, where the half-width of the distribution is modified to
ensure 70\% acceptance. The Cauchy update has changes on all scales,
ensuring that the whole parameter space can be spanned. In both cases,
the general features of the spectrum are recovered. It can be seen
that as few as 100 data points are necessary to determine the
spectrum, although they should be measured as accurately as
possible. In particular, the RMC result shows that the correct form
for the entropy has been used in the maximum entropy
algorithm. Ringing is clearly inherent to both of the techniques shown
here, although RMC performs slightly better. In reality, such a sharp
cutoff in the DOS is not expected, and the spectrum is most likely to
vary continuously to zero, which will remove some of the errors. The
ringing is a limiting factor to the resolution of this technique, and
means that the specific details of the van Hove singularities cannot
be determined.

\begin{figure}
\begin{indented}\item[]
\includegraphics[width=70mm,height=100mm,angle=270]{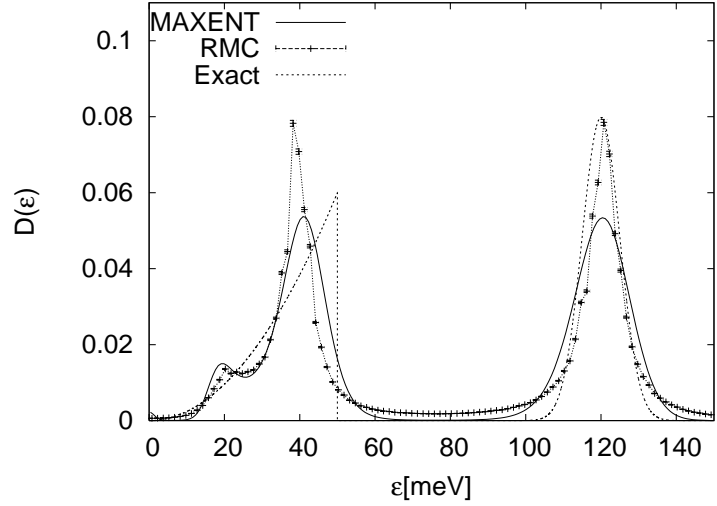}
\end{indented}
\caption{A more realistic simulated phonon DOS with one optical mode
(represented as a Gaussian centred about 120meV with a half-width of
5meV) and one acoustic mode (represented by a Debye model with
$\omega_D=50$meV). Data error is set to 0.1\%. Both Bryan's algorithm
and RMC were used. The method is able to distinguish between optical
and acoustic modes which are on different energy scales without
recourse to alternative models. The current method therefore
represents a major improvement on existing techniques.}
\label{fig:mixed}
\end{figure}

A more realistic simulated phonon DOS representing one optical and one
acoustic mode. Data error is fixed at 0.1\%, and Bryan's algorithm and
RMC are used. The acoustic phonon is represented as a Debye mode with
$\omega_D=50\mathrm{meV}$. The optical mode is modeled as a Gaussian
centred about $120\mathrm{meV}$ with half-width $5\mathrm{meV}$. As
before, there is some ringing associated with the acoustic
mode. However, the method is capable of dealing with densities of
states on all energy scales, which is not possible with existing
methods. As before, it is apparent that the RMC algorithm gives a more
accurate answer. In particular, RMC is much better at determining the
height of the high energy peak. Again, there are some anomalous states
present between the modes and at very high energy, and the RMC clearly
has some error in the continuity of the curves. It would take a very
large but not impractical number of Monte-Carlo measurements to reduce
those errors. With increasing computational power, spectral RMC
calculations will clearly become more feasible.

It is typical to use the specific heat capacity of diamond as a
benchmark for new techniques. To finish, I apply the maximum entropy
and RMC techniques to the data from references
\cite{victor,desorbo}. The MAXENT analysis is very quick, taking only
a few seconds on a modern PC ($\sim 1$GHz). The RMC analysis took a
few hours and is typically run overnight. A data error of 0.5\% is
taken according to reference \cite{victor,desorbo}. For the maximum
entropy algorithm, a flat default model, normalised to unity, and
spread between 0 and 220meV was used.

\begin{figure}
\begin{indented}\item[]
\includegraphics[width=70mm,height=100mm,angle=270]{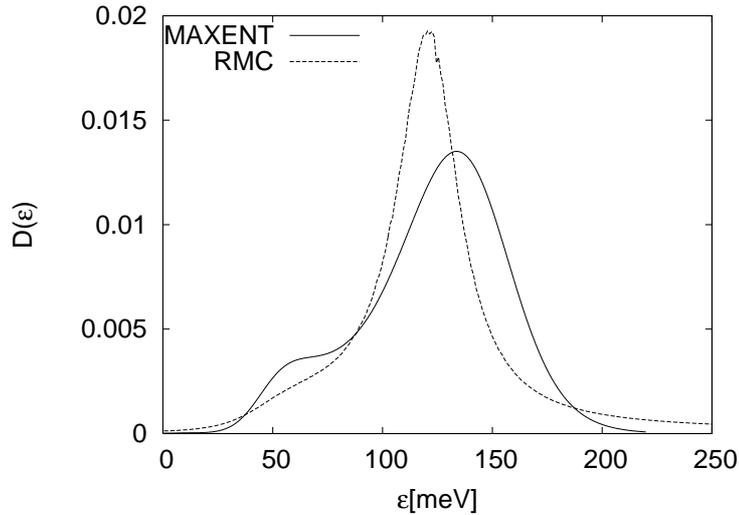}
\end{indented}
\caption{The phonon DOS extracted from the heat capacity of diamond.}
\label{diamond_dos}
\end{figure}

Figure \ref{diamond_dos} shows the results from the fitting. First,
Einstein's expression for the specific heat was fitted to the
available data using Gnuplot's least squares refinement. The energy of
the corresponding Einstein phonon was found to be
$111\pm1.3$meV. Next, the maximum entropy and spectral RMC procedures
were applied. The results show that the general form of the DOS is
very similar to that of Debye, which is reassuring, since the
variation of the measured ``Debye temperature'' with decreasing
temperature is small, indicating a form very similar to the Debye
model \cite{burns}. In order to determine the DOS in higher detail,
either more data points, or higher accuracy data are
required. Typically, neutron scattering studies where the DOS is
obtained indirectly using e.g. a shell model to determine the phonon
dispersion have additional structure. However, it should be noted that
in a typical neutron scattering experiment, the dispersion is measured
along the symmetry directions only, so most of the DOS is obtained by
interpolation and some of the additional structure is spurious.

\section{Conclusion}
\label{sec:conclusions}

I have introduced a general maximum entropy approach for the
computation of the phonon DOS of materials directly from specific heat
measurements. The method is model independent, as opposed to standard
methods of determination such as fitting to Debye or Einstein modes,
or a specific form of the phonon DOS. The Baysian nature of MAXENT
ensures that the best fit to the data is found, without the
over-fitting often associated with least squares approaches. This
method will be of a general benefit to the experimental condensed
matter physics community, since the measurement of the phonon
dispersion using neutron scattering is expensive, and Raman scattering
is limited to zone centre modes. The method is better than existing
heat-capacity methods, because it can deal with phonons on all energy
scales, and can treat acoustic and optical phonons on an equal
footing, and does not suffer from the unphysical negative densities of
states that are reported in reference \cite{loram}.

There are some caveats to this approach. The specific heat at constant
volume ($c_v$) is required. Such measurements are slightly more
difficult to carry out than the specific heat at constant
pressure. However, at moderate temperatures, the specific heat at
constant volume and pressure converge \footnote{For high temperatures
$c_p$ is linear, since $c_v$ saturates to a constant and
$c_p-c_v=\alpha T$ (from thermodynamics - alpha is a constant related
to the compressibility). Therefore it is also possible to determine
$c_v$ directly from $c_p$.}. Details of the conversion between $c_v$
and $c_p$ measurements can be found in reference \cite{victor}. Also,
it should be noted that in materials with very strong electron-phonon
interaction, the phonon DOS is likely to change with temperature. In that
case, the MAXENT derived DOS should be considered to be an
approximation to the true result.

It is for these reasons that the method should not be considered as a
replacement for techniques such as neutron and Raman scattering, but
as a complementary method, either for the quick extraction of
parameters (for example when it is required to separate phonon and
magnetic contributions to the specific heat), when the phonon DOS is
averaged through an integral transform (such as applying the
Eliashberg theory of superconductivity \cite{eliashberg1960a}) or when
it is necessary to have a good idea where to look for excitations
(such as in neutron scattering experiments). It is certainly expected
to be a good way to determine which materials might be interesting
enough for further study.

\ack

I would like to thank Emma Chung for useful discussions and the
Condensed Matter Physics group at the University of Leicester for
hospitality.

\vspace{5mm}

\bibliographystyle{unsrt}
\bibliography{phonondos}

\begin{thebibliography}{10}

\bibitem{einstein1907a}
A.Einstein.
\newblock The planck theory of radiation and the theory of specific heat.
\newblock {\em Ann. d. Phys.}, 22:180, 1907.

\bibitem{ashcroft}
N.W.Ashcroft and N.D.Mermin.
\newblock {\em Solid state physics}.
\newblock Saunders College Publishing, 1976.

\bibitem{chung2004a}
E.M.L.Chung, M.R.Lees, G.J.McIntyre, C.Wilkinson, G.Balakrishnan, J.P.Hague,
  D.Visser, and D.McK.Paul.
\newblock {\em J. Phys: Condens. Matter}, 16:7837, 2004.

\bibitem{lanzara2001a}
A.Lanzara, P.V.Bogdanov, X.J.Zhou, S.A.Kellar, D.L.Feng, E.D.Lu, T.Yoshida,
  H.Eisaki, A.Fujimori, K.Kishio, J.-I.Shimoyama, T.Noda, S.Uchida, Z.Hussa,
  and Z.-X.Shen.
\newblock {\em Nature}, 412:6846, 2001.

\bibitem{loram}
J.W.Loram.
\newblock {\em J. Phys. C: Solid State Phys.}, 19:6113, 1986.

\bibitem{wen}
T.Wen, G.Ma, X.-X.Dai, J.-X.Dai, and W.E.Evenson.
\newblock {\em J. Phys: Condens. Matter}, 15:225, 2003.

\bibitem{montroll}
E.W.Montroll.
\newblock {\em J. Chem. Phys.}, 10:218, 1942.

\bibitem{kroll}
W.Kroll.
\newblock {\em Progr. Theor. Phys. Japan}, 8:457, 1952.

\bibitem{chambers}
R.G.Chambers.
\newblock {\em Proc. Phys. Soc.}, 78:941, 1961.

\bibitem{bryan1989a}
P.Foug\`{e}res, editor.
\newblock {\em Solving oversampled data problems by maximum entropy}, Maximum
  Entropy and Baysian Methods. Kluwer, 1989.

\bibitem{victor}
A.C.Victor.
\newblock {\em J. Chem. Phys.}, 36:1903, 1962.

\bibitem{desorbo}
W.DeSorbo.
\newblock {\em J. Chem. Phys.}, 21:876, 1953.

\bibitem{burns}
G.Burns.
\newblock {\em Solid State Physics}.
\newblock Academic Press, 1985.

\bibitem{eliashberg1960a}
G.M.Eliashberg.
\newblock Interactions between electrons and lattice vibrations in a
  superconductor.
\newblock {\em JETP letters}, 11:696, 1960.

\end{thebibliography}

\end{document}